# Enterprise Portal: from Model to Implementation

Sergey V. Zykov
ITERA Oil and Gas Company
Moscow, Russia
e-mail: szykov@itera.ru

**Abstract**[1]

Portal technology can significantly improve the entire corporate information infrastructure. The approach proposed is based on rigorous and consistent (meta)data model and provides for efficient and accurate front-end integration of heterogeneous corporate applications including enterprise resource planning (ERP) systems, multimedia data warehouses and proprietary content databases. The methodology proposed embraces entire software lifecycle; it is illustrated by an enterprise-level Intranet portal implementation.

## 1. Introduction

Rapidly changing market conditions require fast and flexible enterprise information management based on mission-critical software (comprehensive ERP and huge data warehouse) integration on the Internet technology foundation.

During the recent decades, the data models (DM) and architectures have been changed significantly to support object methodologies and interoperability. Attempts of enterprise application integration have also been undertaken [2,5-7].

The paper presents an integrated data and metadata model, application of the model for integrating heterogeneous corporate databases, warehouses and content repositories, formal approach to building a web portal enterprise-level solution and an overview of improved software implementation. Research methods are based on a creative synthesis of fundamental statements of semantic networks [10], abstract machines [4], categories [1] and lambda calculus [12].

The data model (DM) introduced provides an open, integrated, problem-oriented, event-driven data and metadata management of dynamic, heterogeneous, weak-structured problem domains in a more adequate way than previously known ones. The model allows to produce software architecture and interface solutions for open, distributed, interoperable environments supporting front-end, versatile data warehousing and instant Internet availability on the basis of component, UML, business-process reengineering (BPR) and web portal technologies.

## 2. Architecture and Interface Considerations

Problem domain features require support of dynamic assignment-based comprehensive enterprise activity assessment.

The interface should support dynamic variation of mandatory and optional input fields, flexible changes of access rights and personal preferences as well as consistent data integrity.

Software architecture should provide interoperability, expandability, flexible adjustability to problem domain changes as well as fast and reliable (meta)data updates.

## 3. The Integrated Data and Metadata Model

### 3.1 The Data Object Model

Current problem domain models are not fully adequate to dynamic and static semantics. Moreover, state-of-the-art methods of enterprise applications development do not result in solutions of a wide application range; quite a number of commercial ERP implementations does not provide sufficient flexibility of heterogeneous data handling. According to research results of the problem domain, a computational data model based on object calculus has been built. The model is an innovative synthesis of finite sequence, category, semantic network, and abstract machine theories.

Data objects (DO) of the DM can be represented as follows:

---

DO = < concept, individual, state >,

where a *concept* is understood as a collection of functions with the same definition range and the same value range. An *individual* implies an entity selected by a problem domain expert, who indicates the identifying properties. *State* changes simulate dynamics of problem domain individuals.

Compared to research results known as yet, the proposed DM features more adequate dynamics mapping for heterogeneous problem domains. The DM also benefits better support for problem-oriented integrated data management. In architecture and interface aspects the DM provides straightforward iterative design of open, distributed, interoperable enterprise software based on UML and BPR methodologies. As far as implementation part is concerned, multi-repository information processing of heterogeneous problem domains is supported. Thus, front-end data access is provided which is based on event-driven procedures, dynamic SQL with triggers and web-portal technologies.

The suggested computational DM is based on the two-level conceptualization scheme [14]. *Conceptualization* implies a process of establishing relationship between problem domain concepts.

Individuals h, according to the assigned types T, are united in assignment-dependent collections, thus forming the sort variables:

$H_T(I) = \{h \mid h : I \rightarrow T\}$.

Such formalism is used to simulate problem domain dynamics.

Data model individuals identification requires a unique predicate function $\Phi$, which is dependent on data object d from problem domain D:

$\| Ix\ \Phi(x) \| i = d \Leftrightarrow \{d\} = \{\bar{d} \in D \mid \|\Phi(\bar{d})\| i = 1\}$.

### 3.2 The Metadata Object Model

Let us introduce the compression principle for the computational data object model

$C = Iy: [D]\ x : D(y(x) \leftrightarrow \Phi) = \{x : D \mid \Phi\}$

that allows to apply the model to concepts, individuals and states separately, as well as to data objects as a whole. The suggested computational metadata model expands traditional models, such as ER-model [3], by adding the following metadata compression principle:

$x^{j+1}\ Iz^{j+1}: [\ldots[D]\ldots]\ \forall x^j: [\ldots[D]\ldots]\ (z^{j+1}(x^j)\ \Phi^j)$,

where

$z^{j+1}, x^{j+1}$ – metadata predicate characters for level j,

$x^j$ – individual and $\Phi^j$ – construction of DO definition language for level j.

The integrated model for (meta)data objects and states benefits scalability, expandability, (meta)data encapsulation and transparent visualization by means of semantic networks [10] that can be mapped to UML diagrams. Expandability, adequacy, neutrality and semantic completeness of the model provide problem-oriented software design and (meta)data consistency throughout the implementation process.

Semantics of computational model of objects of data, metadata and states can be adequately and uniformly formalized by means of typed lambda calculus, combinatory logic, abstract machines, and scenarios based on semantic networks.

### 3.3 Model Application for Portal-Based Enterprise Integrated Software Solution

Let us consider a web browser as a universal client-side software. Users alien to huge ERP, proprietary and/or legacy software are usually familiar with it. Web browsers are embedded into most operation systems; they are easily installable and customizable.

A web browser seems to be the lowest common denominator and a user-friendly solution for heterogeneous software integration into an enterprise portal. Moreover, for a huge and geographically spread enterprise, it makes a consistent and uniform interface to retrieve personalized data to virtually any location. A universal web browser interface is equally applicable to Internet, Intranets and Extranets.

To illustrate the model application to enterprise portal implementation, let us consider the following parameters of client appearance and behavior: data access rights, personal preferences (fonts, color settings, etc.), web browser settings (links, cache, history, etc.) and data access device (Web TV, PDA, mobile phone, terminal, etc.) profile.

Let us assume that A and B are sets. Let $B^A$ stand for the mapping from A to B:

$B^A = \{f \mid f : A \rightarrow B\}$.

Let us assign the $\|\circ\|$ evaluation function to $B^A$:

$\|\circ\| = \{f \mid f\ B^A \times A \rightarrow B\}$.

In this case

$\|\circ\| = (<f, x>) = f(x)$,

and, consequently,

$\| <f, x> \| = f(x)$.

Now let us construct the semantics network language model. Let us consider an ordered pair of DO of the form L=<R,C>, where R=$\{R_1, R_2, \ldots\}$ is a set of dyadic predicate symbols and C=$\{C_1, C_2, \ldots\}$ is a set of constants. Therewith, atoms of the model correspond to simple frames in terms of N.D. Roussopulos [10], and expressions denote problem domain individuals. Let us construct a frame evaluation procedure using the $\|\circ\|$ evaluation function.

At this point let us consider an example of a user profile evaluation procedure based on the suggested data model. Let F functional denote the most general class of users (similar to *Everyone* in Microsoft Windows operating systems family). Let the assignment

$s$={high resolution graphics, multimedia}

correspond to user specific settings. Let F($s$) stand for the set of users, for whom the specific settings are restricted to high resolution graphics and multimedia.

Let the assignment

$p$={registered, unregistered, corporate}

correspond to user registration status. Let F($s$)($p$) designate the subset of users with high graphics and multimedia preferences for whom a registration status is assigned, i.e., those who have already visited the enterprise portal.

For the sake of simplicity and without loss of generality, let us consider that portal users set earlier referred to as functional F, is dependent upon browser settings ($v$), data access device type ($e$), personal preferences and access rights:

F=F(($v$), ($e$), …).

In this case,

$||F=F((v), (e), …)||$

means a formal procedure that evaluates the parameterized functional, while

$||F=F((v), (e), …)(s)||$

evaluates users with given specific settings ($s$), and

$||F=F((v), (e), …)(s)(p)||$

evaluates users with given specific settings ($s$) and registration status ($p$). The introduced functional F can be considered an example of computational model for profile evaluation of portal user sets of variable size, from user groups to individuals.

Let us demonstrate that the suggested two-level conceptualization scheme is sufficient for the model adequacy. Let us introduce the following denotations:

$||r_i|| = \{r_{c.s.}, r_{r.s.}\}$ – specific costs;

$||z_i|| = \{z_{c.s.}, z_{r.s.}\}$ – segmentation degree (i.e., possibility of splitting portal users into relatively stable and independent groups);

$||q_i|| = q_i$ – overheads;

$||l_i|| = l_i$ – duration of the request processing stage (download, dynamic form or report creation, etc.);

$||n_i|| = n_i$ – number of request processing stages.

Evaluated values are generalized, i.e., there is no uniqueness of value choice for specific costs and segmentation degree. Generalization level decrease is achieved by considering an assignment point s:

$$||z_i||(||s||) = \begin{cases} ||z_i||(\text{higraph}) = z_{\text{higraph}}, \\ ||z_i||(\text{mmedia}) = z_{\text{mmedia}}; \end{cases}$$

$$||r_i||(||s||) = \begin{cases} ||r_i||(\text{higraph}) = r_{\text{higraph}}, \\ ||r_i||(\text{mmedia}) = r_{\text{mmedia}}. \end{cases}$$

Moreover, further generalization level decrease by means of the second assignment p does not succeed:

$||z_i||(||s||)(||p||) = ||z_i||(||s||);$

$||r_i||(||s||)(||p||) = ||r_i||(||s||).$

Such a result can be explained by the fact that the evaluation procedure is dependent upon portal user responsibilities of the data access rights policy.

However, it is obvious that overheads $q_i$ are dependent both on user-specific settings functions and on registration status, i.e. we should assume that

$||q_i|| = \{q_{i\ \text{higraph}}, q_{i\ \text{mmedia}}\}.$

The statement

$||q_i|| = q_i$

implies that

$q_{i\ \text{higraph}} = q_{i\ \text{mmedia}} = q_i$.

Similarly, the case of a multimedia data warehouse results in a quite different metadata framework including a data type set and a scenario (i.e. script) set to handle the data types. Particularly, multimedia data profile is dependent upon the category of data source (audio record, video record, static image) and it can be divided into sub-categories of photos, logos and catalogues for static images.

The portal data model proposed is equally applicable for problem domains of multimedia data warehouses, ERP systems, and proprietary content databases.

## 4. The Enterprise Portal Implementation

### 4.1 Implementation Scheme Customization

During software design process, ERP specification is transformed from problem domain concepts and relationships to data model entities, then, further, to DBMS scheme (where data objects are manipulated by PL\SQL and metadata ones – by Perl language), and, finally, to target enterprise portal (meta)data warehouse. As a result of problem domain analysis, computational DM and generalized scheme of ERP development [17] have been customized to meet the problem domain requirements.

Let us overview the portal implementation from conceptualization standpoint.

The first conceptualization step (Asg1) is as follows (fig.1). According to the navigation point (i.e. an assignment) chosen by a portal user, a certain portal template type (i.e. a concept) is mapped into the corresponding portal template (i.e. an individual). The portal template (often referred to as a portlet elsewhere) is a heterogeneous document-like object that contains slots (i.e. variables) for title, header, footer, formatted text, static images, video clips, tables (grids) etc. as they would appear in the web page of the portal.

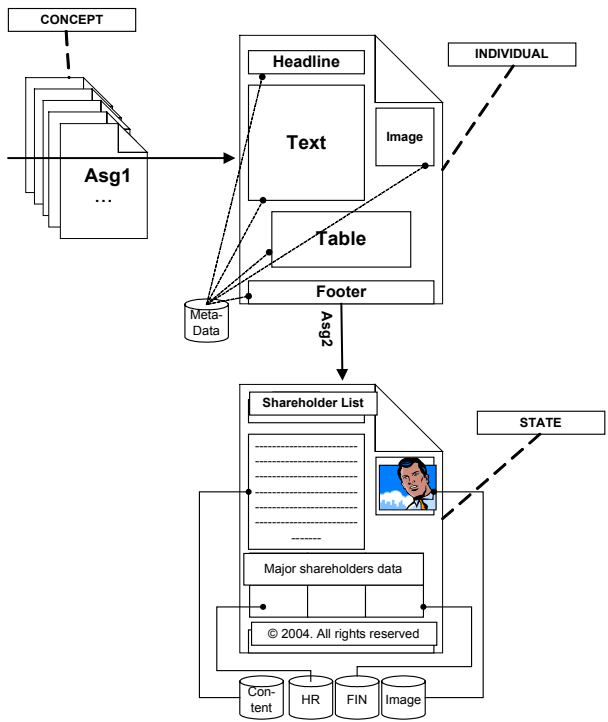

**Figure 1 Enterprise content management scheme based on (meta)data object model**

The second conceptualization step (Asg2) is as follows. According to the query (i.e. assignment) imitated by portal user the slots (i.e. variables) are bound with certain values (web portal page URL, press release title, copyright notice, company president digital image, press release text, table of shareholders) etc. Thus, a web page of the portal is generated dynamically.

Notice that the portal page contains data from heterogeneous sources (e.g., ORACLE Human Resources, ORACLE Financials, multimedia warehouse and content database, etc.). The page also contains metadata (e.g., automatically generated URL). Let us also note that the shareholders table that looks quite homogeneous to the end-user contains data from different locations: human resources and financials databases.

*Content Management System* software based on the DM and *Oracle Portal* act as gateways between heterogeneous data warehouses and corporate Internet and Intranet sites. With such technological enhancements, heterogeneous data sources become integrated. When content-dependent warehouse updates occur in the underlying databases, the portal content is automatically updated (and refreshed) accordingly. Scheduled and manual (meta)data updates and retrievals are also options supported by the implementation.

## 4.2 Event-Driven Architecture and User-Adaptive Interface

According to the enterprise software design sequence [16], a (meta)data processing scheme is suggested that allows users to interact with heterogeneous distributed database in a certain state depending on dynamically activated (i.e., assigned) scripts. Scripts (as data access profiles and stored object-oriented program language procedures) are initiated depending on user-triggered events. Thus, an abstract machine that operates in a domain of states (e.g., the so-called categorical abstract machine [4]) can adequately model content management system.

Scripts provide seamless and intellectual front-end user-to-database connection (a portal user feels cannot recognize the actual number of databases accessed to process the query). Dynamically adjustable database access profiles provide high fault tolerance and data security both for ordinary and privileged portal users in the heterogeneous environment. Depending on semantics-oriented user profile structure, certain database connection and access level profiles are dynamically assigned. The profiles are valid only until the end of data exchange session. According to the role hierarchy, users access data under one of the basic scenario profiles. Access is granted not only to data, but also to metadata (i.e., data object dimensions, integrity constraints, access rights, browser parameters, user preferences, multimedia data types etc.). Naturally, administrative users have extended access to metadata.

Thus, under the model introduced, data and metadata objects are manipulated uniformly. Therewith, software interface is a problem-oriented, intuitive and adjustable one, and it significantly increases performance. Client-side web page object states can change depending on event script execution. Though the warehouse data remains unchanged, user can request an update or make a query. Front-end interface is also client profile dependent. Possible options include personal preferences (multimedia data types, color schemes, screen resolution etc.), data access device and web browser settings.

An essential benefit of Internet orientation of the enterprise portal solution is that corporate users get single access point to business critical data, Intranet information and company Internet site. Let us note that registered (and/or Extranet) users may access some extra data compared to non-privileged ones.

Warehouse data access is also dependent on user profile. At the upper level of data access hierarchy, clients can be divided into administrators, managers and ordinary users. Judging by the profile, data and metadata object states (i.e., system interface) are changed. For instance, web designer rights assignment provides full access to interface elements database, while portal content managers under another assignment get full access to a different warehouse instantiation by means of a different interface.

Scenario-based end-user interface results in higher degree of interactivity, user-friendliness and security. User profiles (i.e. assignments) can be stored in metadata base of visitors, and, depending on their properties, data access and representation levels could be customized. Web pages and their content availability is dependent on user profile. Client-side profile also accounts for layout preferences depending both on user habits and device features, for which the data is customized. User profile and preferences are also vital for visitors/clients analysis; consolidated statistic reports make a solid basis for performance, service and interface optimization.

### 4.3 Implementation Description

The introduced methodology has been practically approved during enterprise information portal implementation in a large fuel-and-energy international group of companies. Attempt has been made to cement the environment containing ERP modules, proprietary multimedia repository and in-house content database with the uniform integrated portal interface.

From the system architecture viewpoint, the enterprise Intranet portal grants variable rights for data input, correction, analysis and output depending on front-end position (i.e., assignment) in user hierarchy. Interactive interface is represented by portal-based problem-oriented form designer, report generator, on-line documentation and administration. The enterprise portal integrated (meta)data warehouse supports integrated storage for data (i.e., information for on-line users) and metadata (i.e., data object dimensions, integrity constraints and business process parameters). During the enterprise portal design, problem domain DM specification (in a semantic network representation) is transformed into use-case UML diagrams, then into ER-scheme by means of versatile CASE-tools and, finally, into the attributes of target databases. *Content Management System* software has been used to transform the ERP modules and warehouse components into a versatile and uniform web portal application.

On the basis of the (meta)data model, an architecture and interface solution for integrated information management software has been developed. To prove adequacy of the model, a script-based software prototype has been designed. Integration of essential (meta)data from ERP components in the enterprise portal had significantly improved implementation.

The portal is capable of accumulating information from problem-oriented modules and third-party warehouses and databases (an example is given in fig.1). For further enhancement of corporate portal performance and efficiency, an event-driven agent for dynamical updates of the data published has been developed.

The enterprise website powered by *Content Management System* has passed a two-year test in a large corporation, while the Intranet enterprise portal solution has passed a half-year test. End-user demand for both of the information resources implemented is increasing constantly.

### 5. Results and Conclusion

A computational DM has been introduced that provides integrated manipulation of data and metadata objects, particularly in rapidly changing heterogeneous problem domains. The model is an alloy of methods of finite sequences, categories, abstract machines and semantic networks.

According to the approach suggested, a comprehensive enterprise web portal interface has been designed. The interface is based on an open and extendable architecture in a form of an abstract machine operating with states. As a first step towards implementing the enterprise resource management solution, a fast event-driven software prototype has been developed on the basis of the designed UML-based interface and architecture scheme. Using the prototype testing results, a pilot version of the full-scale enterprise web portal has been designed. The solution is being customized for corporate information resource management and implemented at a large international corporation.

Enterprise portal solution promises significant decrease in time and costs of implementation. Other major benefits include growth of portability, expandability, scalability and ergonomics in comparison with existing commercial software of the kind. Iterative multi-level software design scheme is based on formal model integrating object-oriented methods of information (data objects) and knowledge (metadata objects) management. Industrial implementation of the Internet-integrated ERP-embracing portal has been carried out using integrated CASE, RAD and portal construction tools. Portal testing experience has proved importance, originality and efficiency of the approach suggested. Theoretical and practical statements outlined in the paper have been approved by successful implementation of the pilot version of full-scale ERP portal solution at a large fuel-and-energy international group of companies. The author is going to continue research in order to turn the initially separated (meta)data components into an integrated corporate Intranet and e-commerce solution.